\documentclass[aps,prb,twocolumn,superscriptaddress]{revtex4-1}
\usepackage{graphicx}
\usepackage{picture}
\usepackage{xcolor}
\usepackage{color}
\usepackage{tikz}
\usepackage{amsmath}
\usepackage{mathrsfs}
\begin{document}

\title{Emission and Absorption Spectrum of Pulse-Driven Two-Level Systems in Dynamic Environments.}

\author{H. F. Fotso}

\affiliation{Department of Physics, University at Albany (SUNY),  Albany, New York 12222, USA}

\begin{abstract}
We study the emission spectrum and absorption spectrum of a quantum emitter when it is driven by various pulse sequences. We consider the Uhrig sequence of nonequidistant $\pi_x$ pulses, the periodic sequence of $\pi_x\pi_y$ pulses and the periodic sequence of $\pi_z$ pulses (phase kicks). We find that, similar to the periodic sequence of $\pi_x$ pulses, the Uhrig sequence of $\pi_x$ pulses has emission and absorption  that are, with small variations, analogous  to those of the resonance fluorescence spectrum. In addition, while the periodic sequence of $\pi_z$ pulses produces a spectrum that is dependent on the detuning between the emitter and the pulse carrier frequency, the Uhrig sequence of nonequidistant $\pi_x$ pulses and the periodic sequence of $\pi_x\pi_y$ pulses have spectra with little dependence on the detuning as long as it stays moderate along with the number of pulses. This implies that they can also, similar to previously studied periodic sequence of $\pi_x$ pulses, be used to tune the emission or absorption of quantum emitters to specific frequencies, to mitigate inhomogeneous broadening and to enhance the production of indistinguishable photons from emitters the solid state.

\end{abstract}

\maketitle

\section{Introduction}

Understanding the dynamics of a quantum emitter coupled to an electromagnetic field, in addition to being a longstanding fundamental question \cite{RF_Heitler_Book1960, Loudon_book1983}, is also of significant current technological interest. In particular, the ability to control the emission or absorption spectrum of a quantum emitter will be of great value to the fields of quantum control, quantum metrology or quantum information processing (QIP) \cite{JeffKimble_qtmInternet, WalmsleyRabitz_PhysToday2003, BraginskyThorne_QtumMeasBook, NielsenChuangBook}.

Indeed, many promising quantum systems that are prime candidates to serve as quantum bits (or qubits) in QIP are quantum emitters in the solid state \cite{Imamoglu_Awschalom_QDOT_PRL_99,ChildressRepeater,HansonAwschalom_QIP_ss_08,
Bernien_Hanson_Nature2013,Pfaff_Hanson_Science2014,Gao_Imamoglu_NatComm2013,Hanson_loopholeFree_Nature2015,Basset_Awschalom_PRL2011,FaraonEtalNatPhot,Sipahigil_SiV,
Rogers_SiV,SantoriVuckovicYamamoto,CarterGammonQDcavity,NV_Review_PhysRep2013}. As such, their emission and absorption spectra are subject to fluctuations in their environment that can lead to the spectrum drifting randomly in time: spectral diffusion \cite{AmbroseMoerner_spectralDiffusion_Nature1991, Fu_Beausoleil_PRL2009}. This phenomenon reduces the efficiency of fundamental operations in QIP such as the photon-mediated entanglement of distant quantum nodes or the coupling of quantum nodes to photonic cavities. It has for this reason received a great deal of attention \cite{Fu_Beausoleil_PRL2009,Bernien_Hanson_Nature2013,Pfaff_Hanson_Science2014,Gao_Imamoglu_NatComm2013,Hanson_loopholeFree_Nature2015,Basset_Awschalom_PRL2011,Hansom_Atature_APL2014,Acosta_Beausoleil_PRL2012, Kuhlmann_Warburton_NatPhys2013,Crooker_Bayer_PRL2010,Matthiesen_Atature_SciRep2014,FotsoEtal_PRL2016, LenhardEtal_PRA2015, TrautmannAlber_PRA2015, YangWrachtrupEtal_NatPhot2016}.

In addition, the absorption spectrum of a quantum emitter such as a Nitrogen-Vacancy (NV) center in diamond can be used to probe weak electromagnetic fields, temperature or forces with very high spatial resolutions. Because of inhomogeneous broadening from the emitters, the sensitivity of such a probe is limited by $1/T_2^*$ instead of $1/T_2$ \cite{ButkerRomalis_AbsorptionMagnetometry, TaylorLukin_BSensing, AcostaBudker_BSensing}.

Recent work has explored the possibility of using adequate pulse sequences to control the spectrum of a quantum emitter \cite{FotsoEtal_PRL2016, JSLee_Khitrin_JPhysB2008, IDS_2017}. In particular, it was shown that a periodic sequence of $\pi_x$ pulses can be used to suppress spectral diffusion from photons produced by a quantum emitter in a diffusion-inducing environment thus enhancing photon indistinguishability \cite{FotsoEtal_PRL2016}. The absorption spectrum of quantum emitters subjected to the same pulse sequence was also studied theoretically \cite{FotsoDobrovitski_Absorption} and experimentally shown to enhance the sensitivity of NV centers used in magnetometry to the $1/T_2$ limit \cite{JoasReinhard_BSensing}. 

These early studies focused on a periodic sequence of $\pi_x$ pulses. The underlying mechanism for modifying the spectral lineshapes is clearly different from that of coherence protection with dynamical decoupling protocols. However, in light of the similarities with dynamical decoupling where different pulse sequences are used with widely varying degrees of success, it is then natural to inquire about the fate of the emission and absorption spectra of quantum emitters if they are driven by other pulse sequences. In this paper, we analyze the emission and absorption spectra of quantum emitters when they are driven by different other pulse sequences. We study the Uhrig sequence \cite{Uhrig_DD_2007} of nonequidistant $\pi_x$ pulses, the periodic sequence of $\pi_x \pi_y$ pulses and the periodic sequence of $\pi_z$ pulses.

We find that the Uhrig sequence of $\pi_x$ pulses has emission and absorption spectra that are analogous  to those of the resonance fluorescence spectrum. In particular, the emission spectrum, has a lineshape analogous to the Mollow triplet of a resonantly driven two-level system \cite{RF_Mollow_PhysRev1969, Autler_Townes_PhysRev1955, KnightMilonni_PhysRep1980} with most of the spectral weight at a central peak flanked with two satellite peaks; similarly, the absorption spectrum is similar to that of two-level system continuously driven on resonance and displays a local maximum at the pulse carrier frequency unlike that of the periodic sequence of $\pi_x$ pulses that has a local minimum at the pulse carrier frequency. The periodic sequence of $\pi_x\pi_y$ pulses has emission and absorption spectra that are qualitatively different from those of $\pi_x$ pulse protocols. They have their main peak at $\pi/2\tau$ and satellite peaks at $-\pi/2\tau$ and $\pi/2\tau + \pi/\tau$, where $\tau$ is the time interval between successive pulses. The periodic sequence of $\pi_z$ pulses has a similar lineshape in both its emission and its absorption spectrum with two peaks of equal spectral weight at $\Delta -\pi/\tau$ and  $\Delta +\pi/\tau$. Where $\Delta$ is the detuning between the emitter and the pulse carrier frequency. The Uhrig sequence of $\pi_x$ pulses and the periodic sequence of $\pi_x\pi_y$ pulses have spectra with little dependence on the detuning of the emitter with respect to the pulse carrier frequency as long as it stays moderate along with the number of pulses. This implies that they can also be used to tailor the emission or absorption of quantum emitters, to mitigate inhomogeneous broadening and to enhance the production of indistinguishable photons from emitters in dynamic environments.

The rest of the paper is organized as follows. In Sec.~\ref{sec:Model}, we describe the model of the quantum emitter coupled to the radiation field and controlled by various pulse sequences, and the master equations governing the system dynamics. In Sec.~\ref{sec:Methods}, we describe the methods used to obtain the emission and absorption spectra. In Sec.~\ref{sec:Results} we present results that show the ability to tune the emission and absorption spectra with appropriate pulse sequences. Sec.~\ref{sec:Conclusions} presents our conclusions.

\begin{figure}
\includegraphics[height=8.50cm, width=8.50cm]{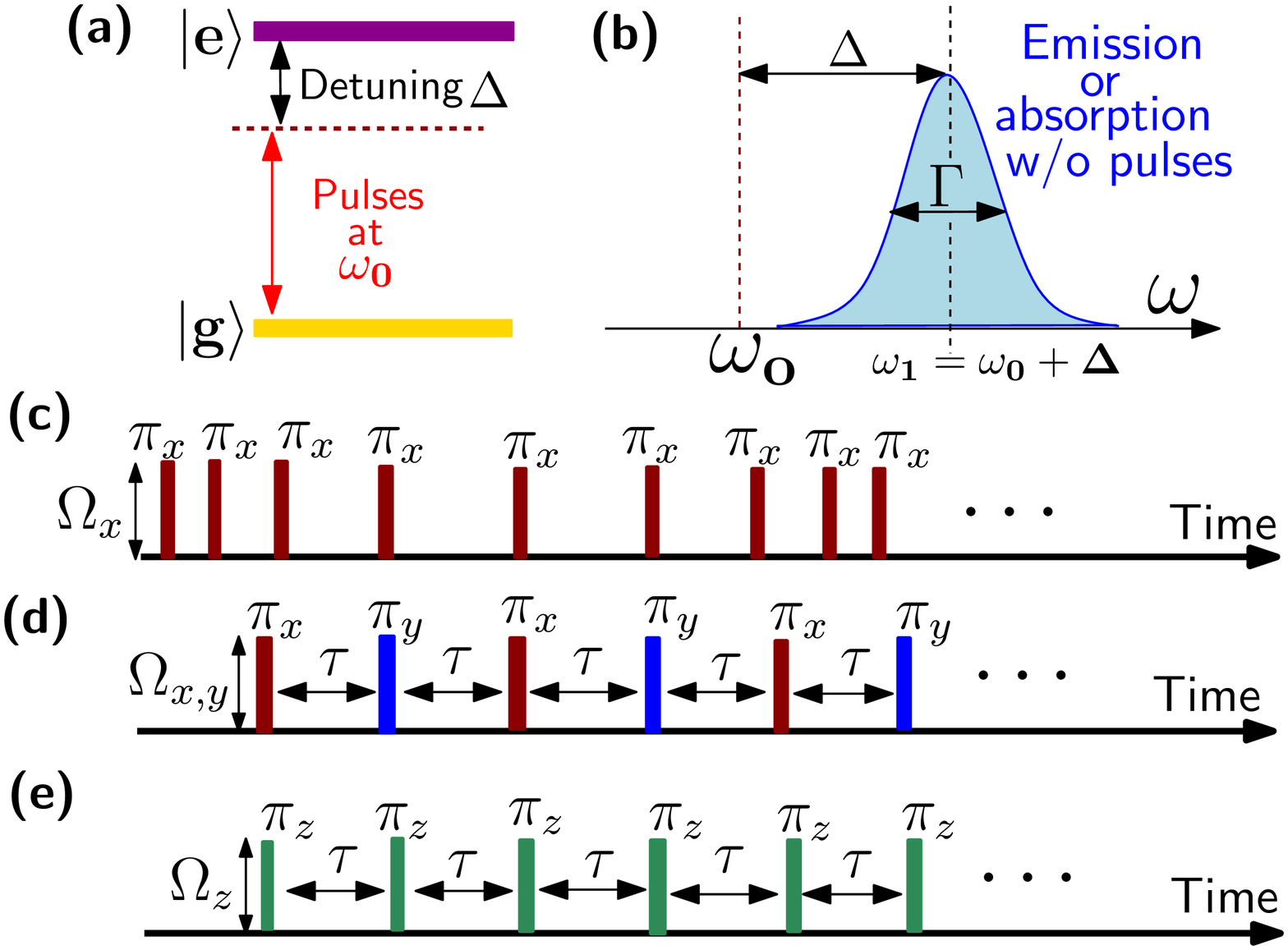}
\caption{(Color online) (a) Schematic representation of a two-level system with ground state $|g\rangle$ and excited state $|e\rangle$ separated by energy $\omega_1 = \omega_0 + \Delta$. (b) In the absence of any driving field, the emission and absorption spectra have a lorentzian lineshape centered around $\omega_1$. (c) Uhrig Sequence of $\pi_x$ pulses. (d) Periodic sequence of $\pi_x\pi_y$ pulses. (e) Periodic sequence of $\pi_z$ pulses.
} 
\label{fig:skecthAbsorption}
\end{figure}

\section{Modeling the driven \textit{emitter + radiation} system}
\label{sec:Model}

The quantum emitter can be modeled as a two-level system (TLS) with ground state $|g\rangle$ and excited state $|e\rangle$, separated in energy by $E_e - E_g = \hbar\omega_1$. Below we set $\hbar=1$.  The TLS is coupled to normal modes of the electromagnetic radiation field, and is, at appropriate times, driven by pulses of the laser field with the Rabi frequency $\Omega$. Initially, at time $t = 0$, the excited state is assumed to be occupied and the ground state to be empty; additionally, all bosonic modes are initially assumed to be empty. The Hamiltonian describing this system can be written as \cite{Cohen_Tannoudji_Book1992}:
\begin{eqnarray}
 H = \frac{\omega_1}{2}\sigma_z + \sum_k \omega_k a_k^{\dagger} a_k &-& i \sum_{k} g_{k} \left( a^{\dagger}_{k} \sigma_- - a_{k} \sigma_+ \right) \nonumber \\
 &-& \vec{d}\cdot \vec{E}_e. 
  \label{eq:hamiltonian_1}
\end{eqnarray}
The first term corresponds to the TLS, the second term to the radiation bath, the third term to the coupling between the TLS and the radiation bath written in the rotating wave approximation (RWA). The last term corresponds to the coupling of the TLS with the external driving field.
$\vec{d}$ is the electric dipole moment of the TLS and $\vec{E}_e$ is the external driving field that is applied at times prescribed by the pulse sequence. It has amplitude $\vec{\mathscr{E}}$ such that the Rabi frequency is $\Omega_i = \vec{d} \cdot \vec{\mathscr{E}_i}$. We have introduced the standard pseudo-spin Pauli operators for the TLS: $\sigma_z = | e \rangle\langle e| - | g \rangle\langle g|$,  $\sigma_+ = |e\rangle \langle g|$ and $\sigma_- = |g\rangle \langle e| = (\sigma_+)^{\dagger} $. $a^{\dagger}_{k}$ and $a_{k}$ are respectively the creation and the annihilation operator for a photon of mode $k$ with the frequency $\omega_k$, and $g_{k}$ is the coupling strength for this mode to the TLS. 

For a given driving sequence, the last term of (\ref{eq:hamiltonian_2}) can be expanded and written in the RWA. For $\pi_x$ rotations, the relevant Hamiltonian is thus, in the frame rotating at frequency $\omega_0$,:

\begin{eqnarray}
H = \sum_{k} \omega_k a^{\dagger}_{k}a_{k} &+& \frac{\Delta}{2} \sigma_z - i \sum_{k} g_{k} \left( a^{\dagger}_{k} \sigma_- - a_{k} \sigma_+ \right) \nonumber \\
&+& \frac{\Omega_x(t)}{2}(\sigma_+ + \sigma_-). 
\label{eq:hamiltonian_2}
\end{eqnarray}
where all energies are now measured with respect to $\omega_0$ and $\Delta=\omega_1-\omega_0$ is the detuning of the TLS's transition frequency from the pulse carrier frequency. The time-dependence $\Omega_x(t)$ is determined by the pulse sequence. We consider pulses such that $\Omega_i(t) = \Omega_i$ during the pulses and zero otherwise. We will assume $\Omega_i$ to be much larger than all other relevant energy scales so that the pulses are essentially instantaneous. Previous studies have shown that imperfect or finite width pulses do not significantly change the spectral lineshapes~\cite{FotsoEtal_PRL2016}. While the last term in (\ref{eq:hamiltonian_1}), written in the form of (\ref{eq:hamiltonian_2}), amounts to treating the incident field as a classical time-dependent field, it can be shown to be equivalent to the treatment of an initially coherent incident field via a unitary transformation \cite{Cohen_Tannoudji_Book1992, RF_Mollow_PhysRev1969, Jeener_Henin_PRA1986}.

In the absence of all control ($\Omega_i(t)\equiv 0$), the system exhibits spontaneous decay, and the corresponding emission rate is $\Gamma = 2\pi\int g_k^2 \; \delta(\omega_k - \Delta) \; dk$; we normalize our energy and time units so that $\Gamma = 2$, and the corresponding spontaneous emission line has a simple Lorentzian shape $1/(\omega^2 + 1)$, with the half-width equal to 1.
 
To characterize the dynamics of the system and obtain the emission and absorption spectrum, we analyze the time evolution of the density matrix of the emitter, which is written as
\begin{eqnarray}
 \rho(t) &=& \rho_{ee}(t) |e\rangle \langle e| + \rho_{eg}(t) |e\rangle \langle g| \nonumber \\
 &+& \rho_{ge}(t) |g \rangle \langle e|  + \rho_{gg}(t) |g\rangle \langle g| \; ,
\label{eq:TrueDensMatr}
\end{eqnarray}
with $\rho_{ge}^* = \rho_{eg}$. The master equations governing the time-evolution of the density matrix operator can be obtained from the Hamiltonian (\ref{eq:hamiltonian_2}) by using the approximation of independent rate of variations whereby we independently add to the time-evolution of each matrix element of $\rho$, terms due to the radiation bath, the incident field and the damping terms responsible for spontaneous emission~\cite{Cohen_Tannoudji_Book1992}. For $\pi_x$ pulses and for the TLS described by the above Hamiltonian (\ref{eq:hamiltonian_2}), the master equations in the rotating-wave approximation are:
\begin{equation}
\label{eq:MasterEquation_1} 
\begin{split}
\dot{\rho}_{ee} &= i \frac{\Omega_x(t)}{2}(\rho_{eg} - \rho_{ge}) - \Gamma \rho_{ee} \; ,   \\
\dot{\rho}_{gg} &= -i \frac{\Omega_x(t)}{2}(\rho_{eg} - \rho_{ge}) + \Gamma \rho_{ee} \; ,  \\
\dot{\rho}_{ge} &= (i\Delta -\frac{\Gamma}{2}) \rho_{ge} -i \frac{\Omega_x(t)}{2}\left(\rho_{ee} - \rho_{gg}\right) \; , \\
\dot{\rho}_{eg} &= (-i\Delta -\frac{\Gamma}{2}) \rho_{eg} +i \frac{\Omega_x(t)}{2}\left(\rho_{ee} - \rho_{gg}\right) .
\end{split}
\end{equation}
Each $\pi_x$ pulse inverts the populations of the excited and ground state and swaps the values of $\rho_{eg}$ and $\rho_{ge}$. 

Similarly, each $\pi_y$ pulse inverts the populations of the excited and ground state and swaps the values of $\rho_{eg}$ with $-\rho_{ge}$ and vice versa. We will also consider $\pi_z$ pulses that are equivalent to $\pi$ phase kicks leaving $\rho_{ee}$ and $\rho_{gg}$ unchanged and replacing $\rho_{eg}$ by -$\rho_{eg}$ and $\rho_{ge}$ by -$\rho_{ge}$. In general, the effect of a $\pi_i$ pulse can be summarized as:
\begin{equation}
 \rho^{(n)}(0^+) = \sigma_i \rho^{(n)}(0^-) \sigma_i.
 \label{eq:pulseEffectPi_i}
\end{equation}
where $\rho^{(n)}(0^-)$ and $\rho^{(n)}(0^+)$ are the density matrices immediately before and immediately after the $n^{th}$ pulse. $\sigma_i$ is the pseudo-spin Pauli matrix in the \textit{i}-direction in which the pulse applies the rotation.

\begin{figure}
\includegraphics[height=8.50cm, width=8.50cm]{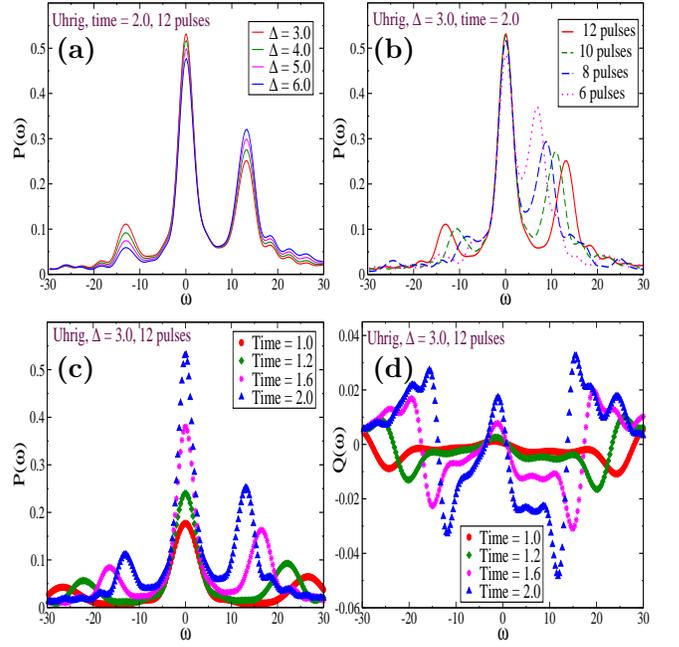} 
\caption{(Color online) Emission and absorption spectra of TLS driven by a Uhrig sequence of $\pi_x$ pulses. (a) Emission for different values of the detuning $\Delta$ between the transition frequency and the pulse carrier frequency ($\Delta = 3.0$ (red),  $\Delta = 4.0$ (green), $\Delta = 5.0$ (magenta), $\Delta = 6.0$ (blue)) after 12 pulses applied during time $T = 2.0$. (b) Emission for fixed detuning ($\Delta = 3.0$) with 6 pulses (doted magenta line), 8 pulses (dashed blue line), 10 pulses (dashed green line), 12 pulses (red line) after time $T = 2.0$. (c) Emission for fixed detuning ($\Delta = 3.0$) with 12 pulses applied during a total time $T = 1.0$ (red circles), $T = 1.2$ (green diamonds), $T = 1.6$ (magenta stars), $T = 2.0$ (blue triangles). (d) Absorption for fixed detuning ($\Delta = 3.0$) with 12 pulses applied during a total time $T = 1.0$ (red circles), $T = 1.2$ (green diamonds), $T = 1.6$ (magenta stars), $T = 2.0$ (blue triangles).
} 
\label{fig:emissionAbsorptionUDD2x2}
\end{figure} 
 
The emission spectrum can be obtained using a narrow-band detector that can be modeled as a two-level absorber with a very sharp transition frequency \cite{Scully_Zubairy_book1997}. The excitation probability of this detector then corresponds to the emission spectrum. It can be expressed as:
\begin{equation}
P(\omega) = A^2 \int_0^T dt \int_0^T ds \langle \sigma_+(t) \sigma_-(s) \rangle exp\left[ -i \omega (t - s) \right],
\label{eq:emissionEq1}
\end{equation}
which can be rewritten as:
\begin{equation}
 P(\omega) = 2 A^2 \mathrm{Re} \int_0^T dt \int_0^{T-t} d\theta \langle \sigma_+(t + \theta) \sigma_-(t) \rangle exp\left[ -i \omega \theta \right].
\end{equation}
On the other hand, the absorption spectrum considered here is measured by determining the energy absorbed from a weak probing field by the TLS while it is simultaneously being driven by the relevant pulse sequence. Since the probing field is assumed to be weak enough that its effects on the populations of the states can be neglected, the absorption spectrum can be calculated within the linear response theory. At long times $T$ the absorption as a function of frequency, $Q(\omega)$, is given by~\cite{Mollow_PhysRevA_3_1972,Mollow_PhysRevA_5_1972}
\begin{eqnarray}
\label{eq:absorptionEq}
Q(\omega) &=& 2 A^2 \\
&\times & \mathrm{Re} \left\{ \int_0^T  dt \int_0^{T-t} d\theta \; \langle \left[ \sigma_-(t) , \sigma_+(t+\theta) \right] \rangle \mathrm{e}^{-i \omega \theta} \right\}, \nonumber
\end{eqnarray}
The angled brackets represent the expectation values. $[O_1, O_2]$ is the commutator of the operators $O_1$ and $O_2$. For the absorption spectrum, the expectation value is evaluated in the absence of the probing field. $\sigma_-(t)$ and  $\sigma_+(t+\theta)$ are the time-dependent operators in the Heisenberg representation, and the expectation values are taken with respect to the initial state of the two-level system (in our case, fully occupied excited state and empty ground state). The constant $A$ is independent of the pulse parameters, does not affect the spectral shape and only affects the absolute scale of the spectrum. The expression (\ref{eq:absorptionEq}) can be rewritten as
\begin{eqnarray}
Q(\omega) &=& 2 A^2 \mathrm{Re} \left\{ \mathcal{P}_2(\omega) - \mathcal{P}_1(\omega) \right\}  \nonumber \\
            &=& P'(\omega) - P(\omega)
\label{eq:absorptionEq2}
\end{eqnarray}
where
\begin{equation}
 \mathcal{P}_2(\omega) =  \int_0^T \; dt\; \int_0^{T-t} d\theta \; \langle  \sigma_-(t) \sigma_+(t+\theta) \rangle \mathrm{e}^{-i \omega \theta}
\end{equation}
and
\begin{equation}
 \mathcal{P}_1(\omega) =  \int_0^T \; dt\; \int_0^{T-t} d\theta \; \langle \sigma_+(t+\theta) \sigma_-(t) \rangle \mathrm{e}^{-i \omega \theta}
\end{equation}
The term $P(\omega)=2 A^2 \mathrm{Re} \left\{ \mathcal{P}_1(\omega) \right\} $ can be recognized to be the emission spectrum of equation (\ref{eq:emissionEq1}). $P'(\omega)=2 A^2 \mathrm{Re} \left\{ \mathcal{P}_2(\omega) \right\}$ can be viewed as the direct absorption so that the difference yields the net absorption~\cite{RF_Mollow_PhysRev1969}. To obtain the absorption spectrum, both terms are evaluated independently before then taking the difference to obtain $Q(\omega)$.

In the absence of any pulses, the emission spectrum and absorption spectrum have Lorentzian-shaped profiles centered at the emitter's frequency that is equal to the detuning $\Delta$ (in the frame rotating at $\omega_0$).

The two-time correlation function $\langle \sigma_+(t+\theta) \sigma_-(t) \rangle$ that appears in the expression for the emission spectrum $P(\omega)$ is usually expressed as a single-time expectation value~\cite{RF_Mollow_PhysRev1969, Scully_Zubairy_book1997, Loudon_book1983} following:
\begin{equation}
\label{eq:expVal_sigmaP_sigmaM}
 \langle \sigma_+(t+\theta) \sigma_-(t) \rangle = \mathrm{Tr} \left[ \rho'(t + \theta) \sigma_+ \right]
\end{equation}
where $\rho^{\prime}(t) = \sigma_- \rho(t) $, and where $\sigma_+$ and $\sigma_-$ are the time-independent operators in the Schr\"odinger picture. $U(t,t')$ is the time-evolution operator for the system described by Eqs.(\ref{eq:MasterEquation_1}). 
Similarly, to evaluate $P'(\omega)$, we rewrite the involved two-time correlation function as:
\begin{equation}
 \label{eq:expVal_sigmaM_sigmaP}
 \langle \sigma_-(t) \sigma_+(t+\theta) \rangle = \mathrm{Tr} \left[ \sigma_+ \rho^{\prime \prime}(t + \theta) \right]
\end{equation}
with $\rho^{\prime \prime}(t) = \rho(t) \sigma_- $.

\begin{figure}
\includegraphics[height=8.50cm, width=8.50cm]{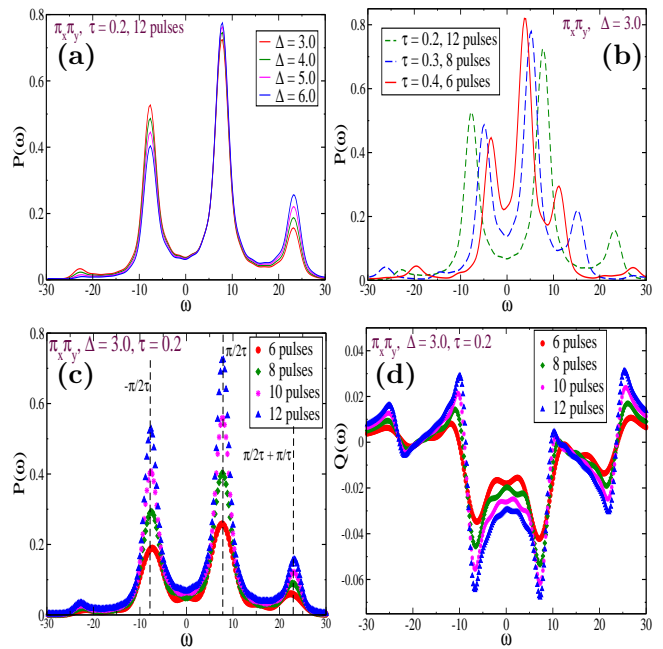}  
\caption{(Color online) Emission and absorption spectra of TLS driven by a periodic sequence of $\pi_x \pi_y$ pulses. (a) Emission for different values of the detuning $\Delta$ between the transition frequency and the pulse carrier frequency ($\Delta = 3.0$ (red),  $\Delta = 4.0$ (green), $\Delta = 5.0$ (magenta), $\Delta = 6.0$ (blue)) after 12 pulses with a time spacing $\tau = 0.2$ between successive pulses. (b) Emission for fixed detuning ($\Delta = 3.0$) with 6 pulses and $\tau =0.4$ (red line), 8 pulses and $\tau = 0.3$ (dashed blue line), 12 pulses and $\tau = 0.2$ (dashed green line). (c) Emission for fixed detuning ($\Delta = 3.0$) and $\tau =0.2$ after 6 pulses (red circles), 8 pulses (green diamonds), 10 pulses (magenta stars), 12 pulses (blue triangles). (d) Absorption for fixed detuning ($\Delta = 3.0$) after 6 pulses (red circles), 8 pulses (green diamonds), 10 pulses (magenta stars), 12 pulses (blue triangles).
} 
\label{fig:emissionAbsorptionXY2x2}
\end{figure}

\section{Analytical and Numerical solutions}
\label{sec:Methods}
The expectation values (\ref{eq:expVal_sigmaP_sigmaM}) and (\ref{eq:expVal_sigmaM_sigmaP}), and the emission and absorption spectra, can be evaluated numerically, or for simple pulse sequences, analytically. Both methods, as demonstrated previously, are in perfect agreement. Alternatively, the Heisenberg equations of motion derived from the Hamiltonian (\ref{eq:hamiltonian_1}) can be integrated analytically within the Markovian approximation. Although this is rather tedious, it also produces, for $\pi_x\pi_y$ and for $\pi_z$ periodic pulse sequences, spectra with lineshape similar to the numerical solution of the master equation with peaks that are identical in their positions and in their relative spectral weights. The analytical solutions provide an insight into the mechanism responsible for the resulting spectrum. Namely, the modulated cancellation of the accumulated phase responsible of the drift of the spectrum to the frequency $\Delta$ (in the frame rotating at frequency $\omega_0$). However, the analytical solution is rather tedious in general. Here, we will present results obtained using our previously demonstrated numerical solution for the different pulse sequences considered. For completeness, we restate the algorithm employed here.

For the numerical solution, the time axis is divided in finite pulse intervals separated by consecutive pulses, and each pulse interval is discretized in smaller steps of length $\Delta t$. Starting at $t = 0$, with the known initial conditions, $\rho_{ee} = 1,\; \rho_{gg} = 0,\; \rho_{eg} = 0,\; \rho_{ge} = 0$, we integrate Eqs.(\ref{eq:MasterEquation_1}) to evolve the matrix elements $\rho_{ee},\; \rho_{eg},\; \rho_{ge},\; \rho_{gg}$ from time $t$ to $t+\Delta t$ and iterate this integration up to the first pulse time $T_1$. We then apply the pulse to the system (i.e Eq.(\ref{eq:pulseEffectPi_i}) to the density matrix operator) before resuming the iterative integration starting from time $T_1$ and up to $T_2$, the time at which the next pulse is applied. This process is repeated until time $T_{N_p}=T$ where $N_p$ is the total number of pulses. It allows us to obtain, $\rho^{\prime}(t)$ and $\rho^{\prime\prime}(t)$ for $t \in [0, T]$. We can then proceed, once again, with the integration of equations(\ref{eq:MasterEquation_1}) starting at $t \in [0, T]$ to obtain $\rho^{\prime}(t+\theta)$ and $\rho^{\prime\prime}(t+ \theta)$ for $\theta \in [0, T-t]$. It is thereon straightforward to obtain $\langle \sigma_+(t+\theta) \sigma_-(t) \rangle$ and $\langle \sigma_-(t) \sigma_+(t+\theta)  \rangle$ from Eqs.(\ref{eq:expVal_sigmaP_sigmaM}) and (\ref{eq:expVal_sigmaM_sigmaP}) respectively. We finally get the emission spectrum and the absorption spectrum by performing the relevant Fourier transforms with respect to $\theta$ and integration over $t$ to get $\mathcal{P}_1(\omega)$ and $\mathcal{P}_2(\omega)$. The real part of $\mathcal{P}_1(\omega)$ gives the emission spectrum and the difference between the real parts of $\mathcal{P}_2(\omega)$ and $\mathcal{P}_1(\omega)$ gives the absorption spectrum.

\section{Results}
\label{sec:Results}

\subsection{Uhrig pulse sequence}
The Uhrig pulse sequence was recently suggested as a sequence of non-equidistant pulses that is optimal in preventing decoherence due to low frequency noise in the environment \cite{Uhrig_DD_2007}. A cycle of the $N^{th}$ order Uhrig pulse sequence is made of $N$ pulses applied at times given by:
\begin{equation}
 T_j = T \mathrm{sin}^2 \frac{j\pi}{2(N+1)},\;\; \mathrm{with } \;\; j = 1, 2, \cdots, N.
\end{equation}
The $(N+1)^{th}$ pulse is applied at $T_{N+1} = T$ and the time sequence is repeated. The Uhrig pulse sequence is illustrated in Fig.\ref{fig:skecthAbsorption}(c). Considering the importance of this pulse sequence in the dynamical decoupling context, it is is useful to study how it affects the emission or absorption spectrum of a quantum emitter in a diffusion-inducing bath.
In Fig.\ref{fig:emissionAbsorptionUDD2x2}, we present results for a quantum emitter driven by a Uhrig sequence of $\pi_x$ pulses. The emission spectrum is found to have a central peak at the pulse carrier frequency, flanked by two satellite peaks the position of which is dependent on the number of pulses. The emission and absorption spectra do not depend much on the detuning $\Delta$ as long as it remains moderate (Fig.\ref{fig:emissionAbsorptionUDD2x2}(a)-(c)). For the same detuning $\Delta$ and fixed time $T$, the central peak is of similar spectral weight and satellite peaks move further away from it as the number of pulses increases (Fig.\ref{fig:emissionAbsorptionUDD2x2}(b)). We note a strong analogy with the spectrum of the resonance fluorescence problem. The absorption spectrum has both positive and negative parts with the former often interpreted as true absorption and the latter as stimulated emission in the direction of the probing field \cite{Wu_et_al_Mollow_PRL1977, BloomMargenau_PhysRev1953, Mollow_PhysRevA_5_1972}. Markedly, the absorption spectrum has a local maximum at the pulse carrier frequency (Fig.\ref{fig:emissionAbsorptionUDD2x2}(d)) unlike that of a periodic sequence of $\pi_x$ pulses that has a local minimum at the pulse carrier frequency.

\begin{figure}
\includegraphics[height=8.0cm, width=8.50cm]{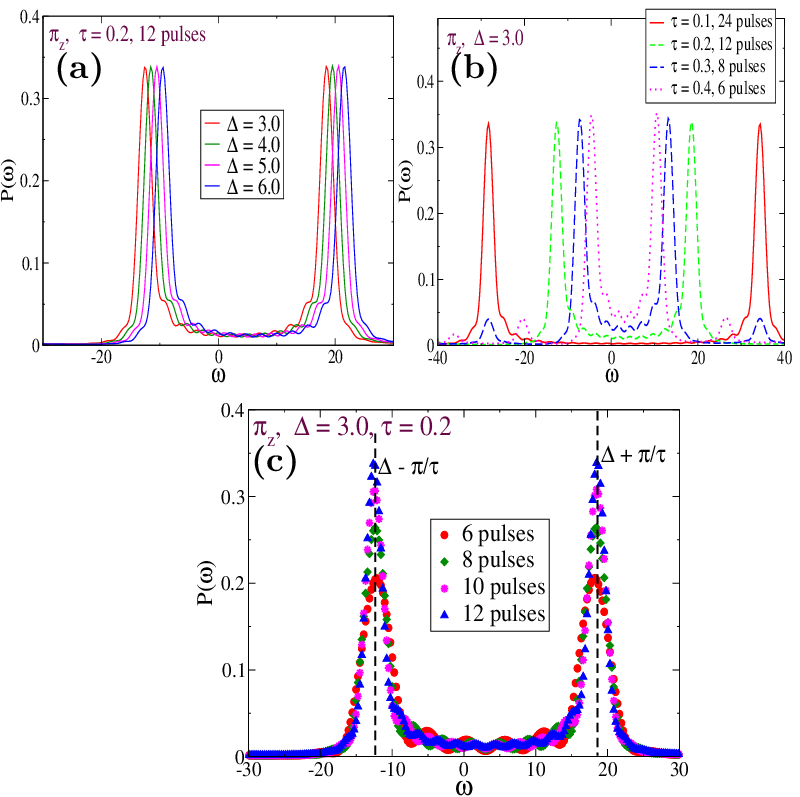} 
\caption{(Color online) Emission spectrum of TLS driven by a periodic sequence of $\pi_z$ pulses (phase kicks). (a) For different values of the detuning $\Delta$ between the transition frequency and the pulse carrier frequency ($\Delta = 3.0$ (red),  $\Delta = 4.0$ (green), $\Delta = 5.0$ (magenta), $\Delta = 6.0$ (blue)) after 12 pulses with a time spacing $\tau = 0.2$ between successive pulses. (b) For fixed detuning ($\Delta = 3.0$) with 6 pulses and $\tau =0.4$ (doted magenta line), 8 pulses and $\tau = 0.3$ (dashed blue line), 12 pulses and $\tau = 0.2$ (dashed green line), 24 pulses and $\tau = 0.1$ (red line). (c) For fixed detuning ($\Delta = 3.0$) and $\tau =0.2$ after 6 pulses (red circles), 8 pulses (green diamonds), 10 pulses (magenta stars), 12 pulses (blue triangles).
} 
\label{fig:emissionPizX3}
\end{figure}

\subsection{$\pi_x \pi_y$ pulse sequence}
In Fig.\ref{fig:emissionAbsorptionXY2x2}, we present the results for an emitter driven by a periodic sequence of $\pi_x\pi_y$ pulses. The emission spectrum has its main peak located at $\pi/2\tau$ with additional peaks at $-\pi/2\tau$ and $\pi/2\tau + \pi/\tau$ (Fig.\ref{fig:emissionAbsorptionXY2x2}(c)) unlike the periodic sequence of $\pi_x$ pulses for which the main peak is at the pulse carrier frequency ($\omega = 0$ in the frame rotating at the pulse carrier frequency). For fixed detuning, the $\tau$-dependence of the emission spectrum is shown in Fig.\ref{fig:emissionAbsorptionXY2x2}(b). The lineshape is established early and the peaks grow in amplitude with time (Fig.\ref{fig:emissionAbsorptionXY2x2}(c)).  The absorption spectrum has its main dip at $\pi/2\tau$ and two additional dips at $-\pi/2\tau$ and $\pi/2\tau + \pi/\tau$ (Fig.\ref{fig:emissionAbsorptionXY2x2}(d)). Both the emission and the absorption spectra show little dependence on $\Delta$ for moderate values of the pulse spacing time $\tau$ as long as $\Delta \lesssim 1/\tau$(Fig.\ref{fig:emissionAbsorptionXY2x2}(a)).


\subsection{$\pi_z$ pulse sequence}
Contrary to the $\pi_x$ and the $\pi_x\pi_y$ pulse sequences, a sequence of $\pi_z$ pulses (or $\pi$ phase kicks), produces a spectrum that depends on the detuning between the pulse carrier frequency and the TLS frequency (\ref{fig:emissionPizX3}(a)). Both the absorption spectrum and the emission spectrum have similar lineshapes. Thus we only present in Fig.\ref{fig:emissionPizX3} the emission spectrum for $\pi_z$ pulses. The emission spectrum is split in two peaks of equal weight located at $\Delta - \pi/\tau$ and $\Delta + \pi/\tau$(\ref{fig:emissionPizX3}(b - c)). Thus, for the same pulse sequence, it depends strongly on the values of $\Delta$ (Fig.\ref{fig:emissionPizX3}(a)). These results are in agreement with those of reference \cite{JSLee_Khitrin_JPhysB2008} where a model of the \textit{atom + radiation} system was solved using direct diagonalization for a finite but large number of bosonic modes.

\section{Conclusions}
\label{sec:Conclusions}

We have studied the emission and absorption spectra of a quantum emitter when it is driven by a pulse sequence. Representing the quantum emitter as a two-level system, we have used the master equation governing the time-evolution of the density matrix operator of the pulse-driven system to obtain the spectrum for the different protocols.  We have considered the case of the Uhrig sequence of $\pi_x$ pulses , a periodic sequence of $\pi_x \pi_y$ pulses and a periodic sequence of $\pi_z$ phase kicks. In the absence  of any driving protocol, the emission and absorption spectra have lorentzian lineshapes centered around the frequency $\Delta = E_e -E_g$ (measured in the frame rotating at the target frequency $\omega_0$). The periodic sequence of $\pi_z$ phase kicks splits the emission spectrum in two peaks  of equal weight  at $\Delta + \pi/\tau$ and $\Delta - \pi/\tau$. It also modifies the absorption spectrum in a similar manner. The Uhrig sequence of $\pi_x$ pulses has an emission and absorption spectrums similar to that of a periodic sequence of $\pi_x$ pulses where a central peak with the bulk of the emission/absorption appears at the pulse carrier frequency with satellite peaks at positive and negative frequencies dependent on the number of pulses. In addition, its absorption spectrum displays a strong analogy with that of the resonance fluorescence with a peak (a maximum) at the pulse carrier frequency as opposed to the minimum that is observed for the periodic sequence of $\pi_x$ pulses. The periodic sequence of $\pi_x \pi_y$ pulses produces an emission spectrum with the main peak located at $\pi/2\tau$ and satellite peaks at $-\pi/2\tau$ and $\pi/2\tau + \pi/\tau$. Similarly, for the Uhrig sequence of $\pi_x$ pulses, the spectra show little dependence on the detuning for a moderate number of pulses over the emission time. These results provide a detailed picture for the emission and absorption spectra of a quantum emitter when it is driven by different pulse sequences. They also indicate that the Uhrig pulse sequence and the periodic sequence of $\pi_x\pi_y$ pulses can be used to control the effect of the environment on the emission and absorption spectrum.

\section*{Acknowledgments} 

We thank V. V. Dobrovitski for inspiring this work and for insightful discussions. We thank V. Mkhitaryan for helpful discussions.

\end{document}